\documentclass[letterpaper,preprint]{revtex4-1}
\usepackage{graphicx}
\usepackage{hyperref}

\begin{document}
\title{KATRIN: an experiment to determine the neutrino mass from the beta decay of tritium}

\author{R.G. Hamish Robertson}

\address{Center for Experimental Nuclear Physics and Astrophysics, University of Washington, Seattle, WA, 98195, USA}

\email{rghr@u.washington.edu}

\author{for the KATRIN Collaboration}

\begin{abstract}
KATRIN is a very large scale tritium-beta-decay experiment to determine the mass of the neutrino.  It is presently under construction at the Karlsruhe Institute of Technology north campus, and makes use of the Karlsruhe Tritium Laboratory built  as a prototype for the ITER project.  The combination of a  large retarding-potential electrostatic-magnetic spectrometer and an intense gaseous molecular tritium source makes possible a sensitivity to neutrino mass of 0.2 eV, about an order of magnitude below present laboratory limits.  The measurement is kinematic and independent of whether the neutrino is Dirac or Majorana.  The status of the project is summarized briefly in this report. 
\end{abstract}
\maketitle
\nopagebreak
\section{Introduction}
In beta decay the electron and neutrino share the available energy in a statistical fashion, and in a small fraction of the decays the electron will take almost all the energy unless some must be reserved for the rest mass of the neutrino.   The phase-space available to the electron near the endpoint is therefore modified by neutrino mass.
In the seven decades of experimental searches for neutrino mass, tritium has been the beta-active nucleus of choice~\cite{Otten:2008zz,Drexlin:2013aaaa}  because it has a low endpoint energy, which makes the modification caused by neutrino mass a larger fraction of the total spectrum:  
\begin{displaymath}
  ^3{\rm H}\,\rightarrow \, ^3{\rm He}^+\,+\,{\rm e}^- \,+\,\overline{\nu}_{\rm e}  \quad {\rm +\ 18580\  eV.}
\end{displaymath}
Atomic or molecular effects are important at the eV level, and T or T$_2$ are simple enough to permit highly precise calculation.  The decay is superallowed, with a short half-life, which reduces the amount of source material needed for a given counting rate and sensitivity. 

Oscillation experiments now define a lower limit to the mass range.  The average mass of the three eigenstates must lie between 20 meV and 1800 meV, the lower limit arising in the `normal' hierarchy if the lightest mass is zero, and the upper limit being set by the Mainz and Troitsk experiments on the beta decay of tritium \cite{Kraus:2004zw,Aseev:2011dq}.   For an electron-flavor-weighted mass $\geq 10$ meV the  beta spectrum is essentially indistinguishable from what  a single massive `electron neutrino'  would produce if mass eigenstates were also flavor eigenstates.     The dependence of the spectral shape on mass is given
by a phase space factor only. Moreover, the 
mass measured is independent of whether the
neutrino is a Majorana or a Dirac particle.   A neutrino mass in the quasi-degenerate regime $\geq 200$ meV that KATRIN is designed to explore would be of cosmological importance, having a substantial influence on the formation of large-scale structure in the universe~\cite{SejersenRiis:2011sj,Kristiansen:2007di,Host:2007wh}.  In addition, KATRIN is sensitive to admixtures of putative sterile neutrinos with masses in the range eV to keV~\cite{Esmaili:2012vg,Formaggio:2011jg,Riis:2010zm}, to right-handed currents~\cite{Bonn:2007su} or new interactions~\cite{Ignatiev:2005nu}, and offers thousand-fold improved sensitivity to the capture of relic neutrinos~\cite{Faessler:2013jla}.

\section{The KATRIN Experiment}

\begin{figure}
\begin{center}
\includegraphics[width=6in]{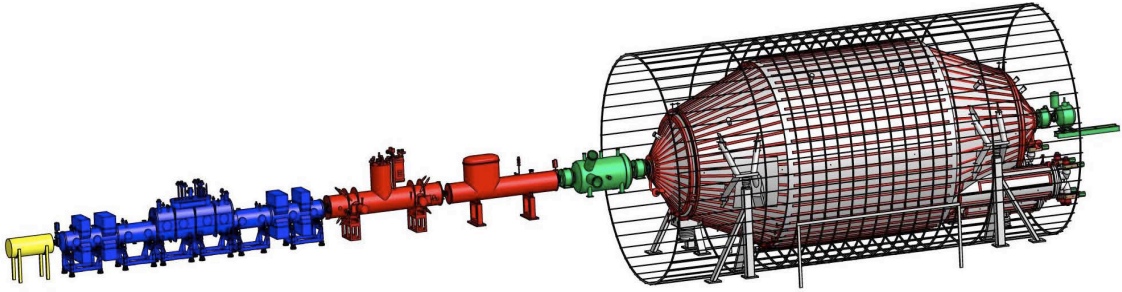}
\end{center}
\caption{\label{fig:overview}Layout of the KATRIN Experiment}
\end{figure}

The KArlsruhe TRItium Neutrino experiment \cite{Angrik:2005ep} consists of seven major subsystems (see Fig. \ref{fig:overview}), a gaseous tritium source, the tritium processing and recirculation system, a differential pumping section, a cryogenic Ar frost pumping section, a pre-spectrometer, a main spectrometer, and the detector system.  

\section{Source and Tritium Recirculation}
The gaseous molecular source consists of a tube 9 cm in diameter and 10 m in length maintained at the desired temperature (27 K) by circulation of two-phase Ne~\cite{Babutzka:2012xd}.  An axial magnetic field of 3.6 T guides electrons toward the spectrometers.  The cryogenic system has been constructed by Research Instruments to stringent thermal performance specifications, including temperature regulation to $\pm30$ mK, and in fact achieved $\pm4$ mK~\cite{Grohmann2009413,Grohmann:2011zz,Grohmann20135}.  The solenoids are being provided by Bruker Advanced Superconductors.  KATRIN is made possible by the existence of TLK, the Karlsruhe Tritium Laboratory, which is the prototype system for the International Thermonuclear Experimental Reactor ITER.  As much as 10 kg/y of isotopically and chemically purified tritium can be circulated in closed loops~\cite{Fischer:2012xs}.
\section{Differential Pumping Section}
Further reduction of tritium pressure is achieved in this section, which consists of a tube with a chicane in a set of superconducting warm-bore solenoids being manufactured by Cryomagnetics Inc.~\cite{Lukic:2011fw}.  
\section{Cryogenic Pumping Section}
Any remaining tritium that escapes through the differential pumping section is trapped in Ar frost, which forms a highly efficient, large-area, chemically inert, and radiation-immune surface.  The CPS is nearing completion at Ansaldo Superconduttori~\cite{Eichelhardt:2011zz}.
\section{Pre-spectrometer}
There are two spectrometers in tandem in KATRIN,  both of the retarding-potential type~\cite{Thummler:2009rz}. The pre-spectrometer operates at a cutoff potential typically 100 eV below the endpoint, preventing most electrons from reaching the main spectrometer~\cite{Prall:2012rx}.  Electrons can ionize residual gas molecules and create slow electrons that are indistinguishable from the signal~\cite{Beck:2009ki}. Reducing the electron flux into the main spectrometer is expected to improve the background near the endpoint substantially.  The pre-spectrometer was the first KATRIN subassembly to be completed, in order to permit some key concepts to be tested before other design elements were frozen.  Two important discoveries were made, one the existence of a parasitic Penning trap~\cite{Formaggio:2012aaaa,Glueck:2011aaaa}  near the ends, which made it impossible to apply high voltage and a magnetic field simultaneously without breakdown.  The addition of appropriate  electrodes suppressed this discharge completely, and the spectrometer now runs uneventfully at  35 kV and 4 T.   A second observation was the presence of radon, which led to long periods of elevated slow-electron production while an energetic beta spiraled back and forth, gradually losing energy~\cite{Frankle:2011xy,Mertens:2012vs}.   Liquid-nitrogen baffles were specified for the main spectrometer to trap radon, the origin of which turned out to be non-evaporable getter (NEG) strips.  
\section{Main Spectrometer}
The main spectrometer is a large stainless-steel vessel 10 m in diameter and 24 m in length~\cite{Osipowicz:2012cj}.  The 200-tonne chamber was fabricated in Deggendorf by MAN-DWE and shipped via  a long water route to Karlsruhe.  It was placed in its building November 29,  2006.  Completion of the interior grid structure, installation of 1000 m of NEG strips, and installation of liquid-nitrogen cooled baffles took place in 2012, and commissioning of the device began in May, 2013.  
\section{Detector}
Electrons surmounting the potential barriers in the spectrometers enter a high axial magnetic field region and are detected in a monolithic 148-pixel Si PIN diode array 10 cm in diameter~\cite{VanDevender:2012rx}.  Two superconducting solenoids can produce up to 6 T, defining the electron beam diameter and the maximum pitch angle accepted from the source.  The magnets were made by Cryomagnetics, Inc. and the detector by Canberra.  Between the solenoids is a region in which various calibration devices -- gamma sources and a photoemissive electron gun -- can be inserted.  Selection of materials, shielding,  and an active veto have been adopted to keep non-beam-associated backgrounds sufficiently low.  
\section{Summary}
The KATRIN experiment is scheduled for initial data-taking in 2015.   Delays have been experienced owing in part to vendors changing hands during the performance of contracts, but these issues are believed now to be completely resolved.  Funding for the project is being provided by the Helmholtz Gemeinschaft, the Bundesministerium f\"{u}r Bildung und Forschung, and the US Department of Energy Office of Nuclear Physics. 

\section*{References}
\bibliographystyle{iopart-num}
\bibliography{testbib}{}

\end{document}